\documentclass[aps,twocolumn,prl,superscriptaddress,amsmath,amssymb]{revtex4-1} %% for PRL
\usepackage{dcolumn}
\usepackage{bm}
\usepackage{amsmath}
\usepackage{txfonts}
\usepackage[T1]{fontenc}
\usepackage{xspace}
\usepackage{ulem}
\usepackage{comment}
\newcommand{\means}[1]{\langle#1\rangle}

\ifx\pdfoutput\undefined
\usepackage[dvipdfmx]{graphicx}
\usepackage[dvipdfmx]{hyperref}
\usepackage[dvipdfmx]{color}
\usepackage[dvipdfmx]{xcolor}
\else
\usepackage{graphicx}
\usepackage{hyperref}
\usepackage{color}
\usepackage{xcolor}
\fi
\begin{document}
\let\emph\textit

\title{
Thermodynamic and Transport Properties in Disordered Kitaev Models
  }
\author{Joji Nasu}
\affiliation{
  Department of Physics, Yokohama National University, Hodogaya, Yokohama 240-8501, Japan
}
\author{Yukitoshi Motome}
\affiliation{
  Department of Applied Physics, University of Tokyo,
  Bunkyo, Tokyo 113-8656, Japan
}

\date{\today}
\begin{abstract}
  Effects of bond randomness and site dilution are systematically investigated for the Kitaev model describing a quantum spin liquid with fractional excitations of itinerant Majorana fermions and localized fluxes.
  We find that, in the high-temperature region where the itinerant Majorana fermions release their entropy, both types of disorders suppress the longitudinal thermal conductivity while keeping the specific heat almost unchanged. 
  This suggests that both disorders reduce the mean-free path of the Majorana fermions.
  On the other hand, in the low-temperature region, the other specific heat peak associated with the entropy release from the localized fluxes is suppressed for both cases, but it is broadened and shifted to the lower-temperature side by the bond randomness, while the position and the width are almost unchanged against the site dilution.
  Contrasting behavior is also found in the thermal Hall effect under a magnetic field; the half quantization of the thermal Hall conductivity is fragile against the site dilution, while it remains for the bond randomness despite the reduced onset temperature.
  We discuss the contrasting behavior from the stability of the topological nature by calculating flux condensation and Majorana excitation gap.
\end{abstract}
\maketitle

%%%%%%%%%%%
% Introduction

Among a lot of research on quantum spin liquids (QSLs), which are quantum states without any conventional magnetic orderings, the Kitaev's seminal work on a localized spin Hamiltonian with exact QSL ground states has brought breakthrough innovations, not only in the field of magnetism but also for quantum information~\cite{kitaev2003fault,Kitaev2006,RevModPhys.87.1,Trebst2017pre,Hermanns2018rev,Knolle2019rev,takagi2019rev,Motome2020rev}.
A promising realization of the model was suggested for transition metal compounds with the strong spin-orbit coupling~\cite{PhysRevLett.102.017205}, which has triggered intensive experimental and theoretical studies over the decade.
In particular, the layered honeycomb compounds, such as $A_2$IrO$_3$ ($A=$Li or Na)~\cite{PhysRevLett.105.027204,PhysRevB.82.064412,PhysRevLett.108.127203,PhysRevB.88.035107,PhysRevLett.110.097204,1367-2630-16-1-013056,PhysRevLett.113.107201,Winter2016} and $\alpha$-RuCl$_3$~\cite{PhysRevB.90.041112,PhysRevB.91.094422,PhysRevLett.114.147201,Johnson2015,PhysRevB.91.144420,Cao20016,yadav2016kitaev,Winter2016,PhysRevB.93.155143,Koitzsch2016}, have been studied as the prime candidates for the Kitaev QSL. 
While these compounds exhibit magnetic orderings at low temperature, unconventional behaviors have been reported above the N\'eel temperature~\cite{PhysRevLett.114.147201,Nasu2016nphys,Do2017majorana,banerjee2016proximate,hirobe2017,Hentrich2018,Hentrich2019} or in an applied magnetic field~\cite{PhysRevB.91.094422,Johnson2015,Sears2017,Wolter2017,janvsa2018observation,Widmann2019,banerjee2018excitations,Baek2017,Zheng2017,Nagai2020}, as the signatures of fractional excitations: 
itinerant Majorana fermions and localized fluxes~\cite{PhysRevB.92.115122,PhysRevLett.112.207203,PhysRevLett.113.187201,winter2017breakdown,Song2016,Nasu2016nphys,Halasz2016,yoshitake2016,Nasu2017,Yoshitake2017PRBb,Yoshitake2017PRBa,Udagawa2018,Yoshitake2020}.
Amongst others, considerable attention has been attracted for the discovery of the half-quantized thermal Hall effect in $\alpha$-RuCl$_3$~\cite{Kasahara2018,kasahara2018majorana,Yokoi2020pre}, as convincing evidence of the chiral Majorana edge mode and non-Abelian anyons~\cite{Kitaev2006}.

%Meanwhile, f
For further unveiling the intrinsic nature of QSLs, it is important to clarify the effects of disorders that inevitably exist in real materials. 
Such disorder effects were experimentally studied, e.g., for solid solutions (Na$_{1-x}$Li$_x$)$_2$IrO$_3$~\cite{Cao2013,Manni2014,Manni2014a,Rolfs2015,gupta2016raman,Hermann2017,Simutis2018}, and the results were theoretically discussed as the effects of bond randomness in the Kitaev model~\cite{Chua2011,Zschocke2015,Andrade2014}.
In a more recent candidate H$_3$LiIr$_2$O$_3$, which does not show any magnetic ordering down to the lowest temperature~\cite{Kitagawa2018nature}, the role of stacking fault or bond randomness due to fluctuations of hydrogen positions was discussed~\cite{Yadav2018,LiYing2018,Knolle2019disorder,Wang2018pre,Geirhos2020pre}.
On the other hand, the replacement of the magnetic ions by nonmagnetic ones has been investigated to clarify the effect of site dilution, e.g., in solid solutions $A_2$(Ir$_{1-x}$Ti$_x$)O$_3$ with $A$=Na and Li~\cite{Manni2014a} and (Ru$_{1-x}$Ir$_x$)Cl$_3$~\cite{Lampen-Kelley2017,Do2018,Do2020}.
Theoretically, it was shown that vacancies and dislocations induce local Majorana zero modes in the Kitaev model~\cite{Willans2010,Willans2011,Santhosh2012,Petrova2013,Petrova2014,Sreejith2016,brennan2016lattice}, manifested in locally-induced magnetic moments and dynamical spin fluctuations~\cite{Udagawa2018,Otten2019}.
However, comprehensive understanding of the disorder effects has not been reached yet. 
In particular, less is known for thermodynamic and transport properties despite their importance for the identification of the pristine nature of the Kitaev QSLs in experiments.

In this Letter, we study the effects of bond randomness and site dilution in the Kitaev model using unbiased quantum Monte Carlo simulations.
In the high-temperature region, we find that the longitudinal thermal conductivity is strongly suppressed by both types of disorder, while the specific heat peak is almost unchanged. 
This suggests that both disorders suppress the mean-free path of the heat carriers, in this case, the itinerant Majorana fermions. 
On the other hand, we show that the two types of disorders work quite differently in the low-temperature region. 
The other peak in the specific heat is smeared and shifted to the lower-temperature side by introducing the bond randomness, but for the site dilution, the position and the width are hardly changed despite the reduced intensity.
We also find that the half-quantization plateau in the thermal Hall conductivity is fragile against the site dilution but tenacious against the bond randomness. 
We discuss the contrasting effects on the topological nature by calculating the flux condensation and the Majorana excitation gap.

%%%%%%%%%%%
% Model

\begin{figure}[t]
  \begin{center}
    \includegraphics[width=\columnwidth,clip]{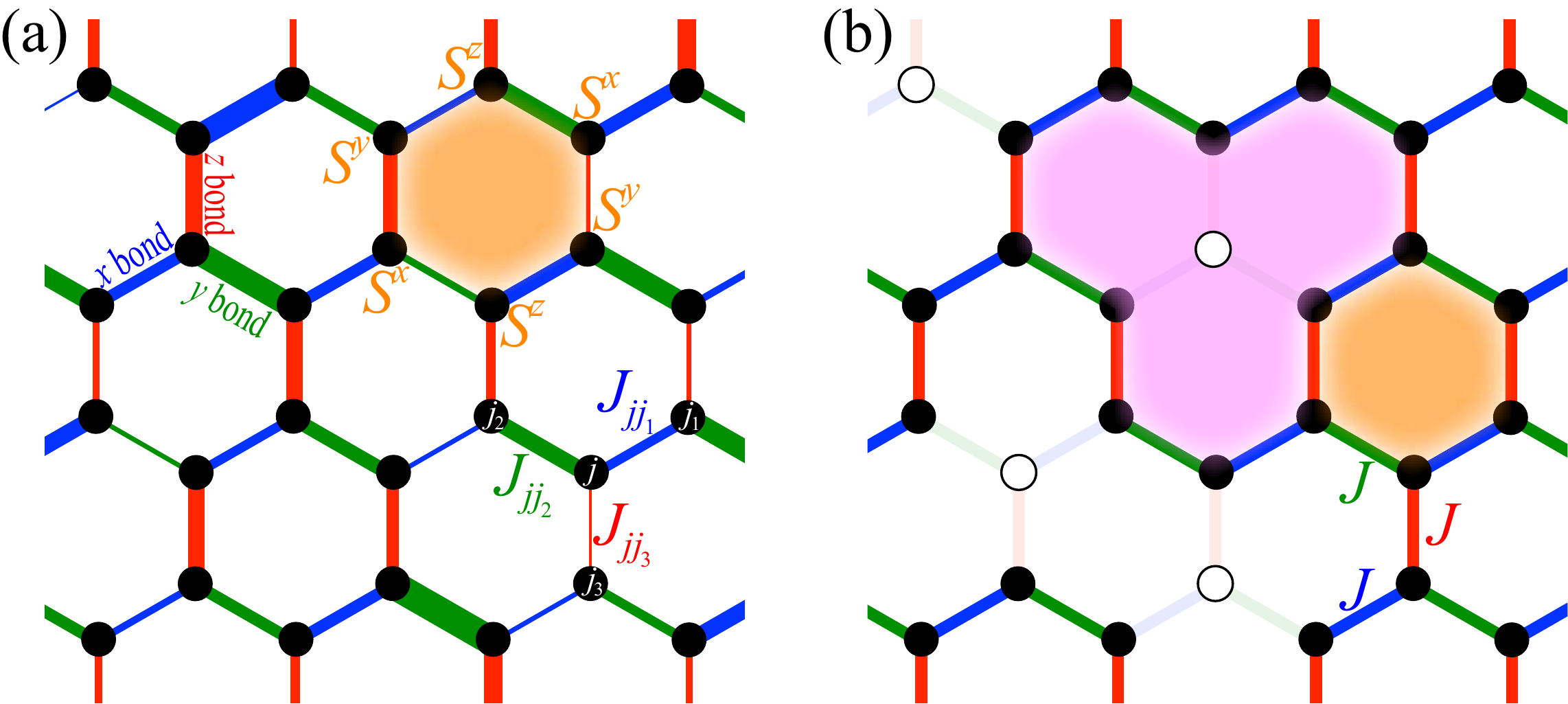}
    \caption{
    (a) Schematic pictures of the Kitaev model with (a) bond randomness and (b) site dilution.
The blue, green, and red bonds represent the $x$, $y$, and $z$ bonds in Eq.~(\ref{eq:Kitaev}), respectively, and their thickness stand for the strength of the coupling constant $J_{jj'}$; the white circles in (b) represent the vacancies. 
The orange hexagons represent fluxes, each composed of the surrounding six spins, while the pink one in (b) is a flux consisting of three hexagons around a vacancy.
  }
    \label{fig_lattice}
  \end{center}
  \end{figure}

To address the disorder effects on the Kitaev QSL, we consider the Kitaev model whose Hamiltonian is given by~\cite{Kitaev2006}
\begin{equation}
  {\cal H}=-\sum_{\means{ij}_\gamma} J_{jj'}S_j^\gamma S_{j'}^\gamma,\label{eq:Kitaev}
\end{equation}
where $S_j^\gamma$ is the $\gamma(=x,y,z)$ component of the spin-$1/2$ operator at site $j$ on a honeycomb lattice with three kinds of nearest-neighbor bonds $\means{jj'}_\gamma$ as shown in Fig.~\ref{fig_lattice}.
We introduce the two types of disorders, the bond randomness and site dilution, separately in Eq.~(\ref{eq:Kitaev}).
For the former, the exchange constant $J_{jj'}$ is generated from a uniform random number in the range of $[J-\zeta,J+\zeta]$, as schematically shown in Fig.~\ref{fig_lattice}(a).
In the following calculations, we consider the range of $0\leq\zeta/J\leq 1$.
On the other hand, for the latter, we consider the situation where some of the spins are randomly removed from the lattice, and accordingly, $J_{jj'}$ connected to the vacancies are set to zero, while the rest are taken to be uniform as $J$, as shown in Fig.~\ref{fig_lattice}(b).
We denote the density of the vacancies by $\rho$.

We study the thermodynamic and transport properties of the disordered models by using a quantum Monte Carlo simulation in the Majorana fermion representation~\cite{PhysRevLett.113.197205,PhysRevB.92.115122}. 
For a given random configuration of $\{J_{jj'}\}$, we measure physical quantities for 100 samples among 20000 MC steps (every 200 steps) after 10000 MC steps for thermalization.
The calculations are performed for the 288-site cluster ($L=12$ and $N=2L^2$) including vacancies under the shifted boundary condition (see Supplemental Material in Ref.~\cite{PhysRevLett.113.197205});  
we note that the finite-site effect is negligible for $N=288$~\cite{Nasu2017}.
The results with statistical errors are evaluated for 20 (10) configurations of $\{J_{jj'}\}$ for the case of the bond randomness (site dilution).

\begin{figure}[t]
\begin{center}
  \includegraphics[width=\columnwidth,clip]{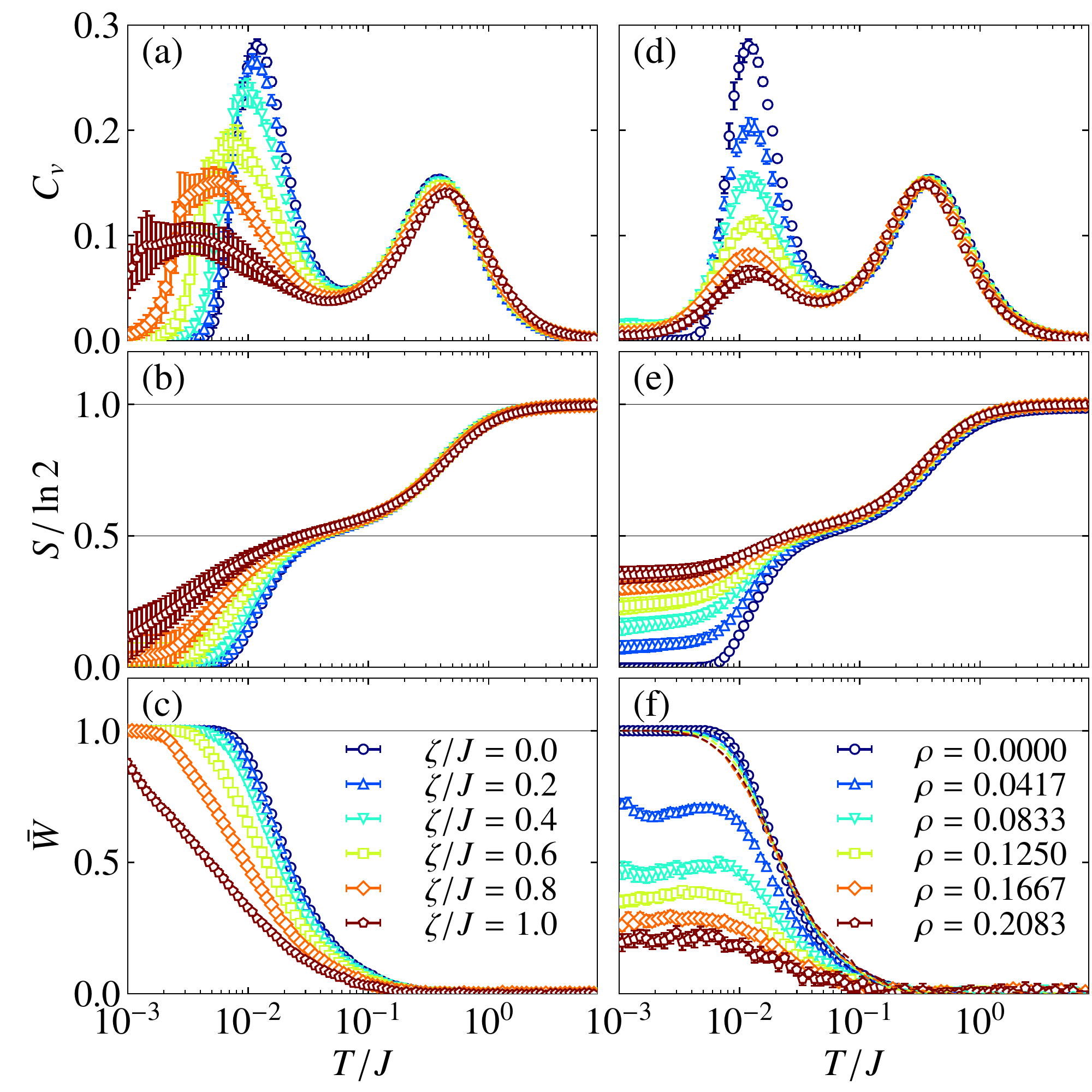}
  \caption{
  (a)--(c) Temperature dependences of (a) the specific heat per spin, (b) the entropy per spin divided by $\ln 2$, and (c) the flux density $\bar{W}$ (see the text for the definition) for the systems with bond randomness.
  (d)--(f) Corresponding results for the systems with site dilution.
  The dashed lines in (f) represent the flux density averaged only for six-site hexagons.
}
  \label{fig_cv}
\end{center}
\end{figure}

Figures~\ref{fig_cv}(a) and \ref{fig_cv}(b) show the temperature dependences of the specific heat per spin, $C_v$, for the bond randomness and site dilution, respectively.
In the pristine case ($\zeta=0$ and $\rho=0$), $C_v$ shows two peaks at $T_L\sim 0.012J$ and $T_H\sim 0.38J$, at each of which half of the entropy, $\frac12 \ln2$, is released as shown in Figs.~\ref{fig_cv}(b) and \ref{fig_cv}(e).
This is a consequence of the thermal fractionalization of quantum spins into itinerant Majorana fermions and localized fluxes~\cite{PhysRevB.92.115122}.
When introducing the disorders, the high-temperature peak is almost unchanged for both cases.
This indicates that both types of disorders do not disturb the entropy release from the itinerant Majorana fermions. 
In contrast, the low-temperature peak is significantly suppressed by the disorders, and surprisingly, exhibits contrasting responses to the two types of disorders:
The peak is smeared and shifted to the low-temperature side for the bond randomness, whereas the peak position and the width are almost unchanged by the site dilution.

The contrasting behaviors are also seen in the entropy.
By the introduction of the bond randomness, the change of the entropy around $T_L$ becomes slow but it approaches zero in the low-temperature limit, as shown in Fig.~\ref{fig_cv}(b).
This is consistent with the behavior of the flux density defined by $\bar{W}=\frac{1}{L^2}\sum_p \means{ W_p}$, where the local conserved quantity $W_p$ is given by $W_p=\prod_{i\in p} 2S_{i}^{\gamma_i}$ for each hexagonal plaquette $p$ with $\gamma_i$ being the bond component not belonging to the edges of $p$ at site $i$~\cite{Kitaev2006} [see Fig.~\ref{fig_lattice}(a)]:
As shown in Fig.~\ref{fig_cv}(c), $\bar{W}$ is suppressed by the bond randomness but approaches unity with decreasing temperature, suggesting that the flux-free ground state with all $W_p=+1$ is reached.
The results indicate that the entropy associated with the localized fluxes are fully released even in the presence of the bond randomness in this range of $\zeta/J$. 
We note that this is consistent with the previous work for the ground state that predicts a transition from flux-free to random-flux states at $\zeta_c/J\simeq 0.96$~\cite{Zschocke2015}.

\begin{figure}[t]
  \begin{center}
    \includegraphics[width=\columnwidth,clip]{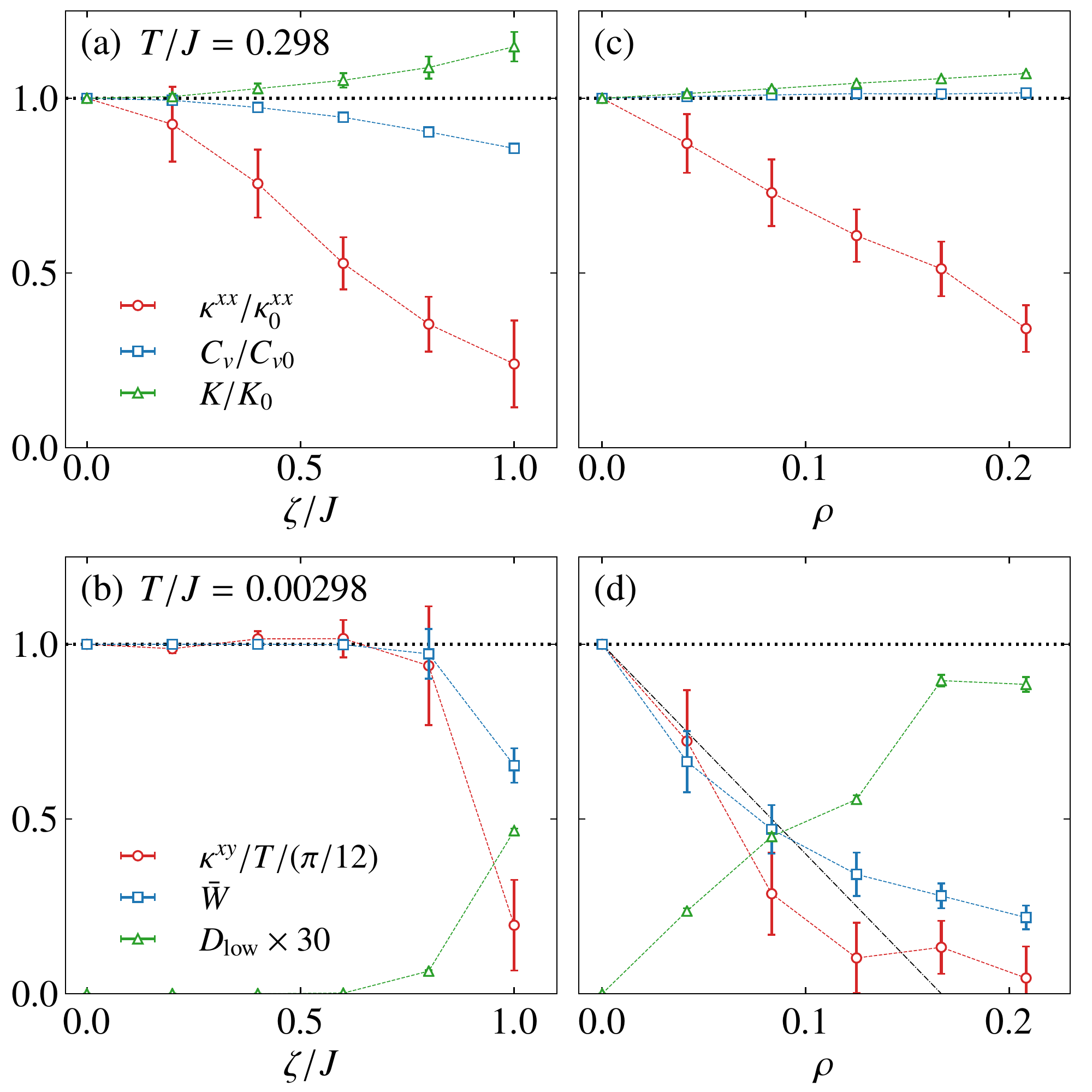}
    \caption{
      (a) Comparison between $\kappa^{xx}$, $C_v$, and $K$ as functions of the bond disorder strength $\zeta$ at $T/J=0.
298$.
      The data are normalized by the values in the pristine case ($\zeta=0$). 
      (b) Comparison between $\kappa^{xy}/T/(\pi/12)$, $\bar{W}$, and $D_{\rm low}$ in the presence of the effective magnetic field $\tilde{h}/J=0.06$ at $T/J=0.00298$.
      (c) and (d)  Corresponding results for the systems with site dilution.
      In (d), the dashed-dotted line represents $1-6\rho$.
    }
    \label{fig_disorder}
  \end{center}
  \end{figure}

On the other hand, in the case of the site dilution, the entropy does not vanish at the lowest temperature calculated here, as shown in Fig.~\ref{fig_cv}(e).
This is attributed to the flux fluctuations in larger plaquettes generated by vacancies [see Fig.~\ref{fig_lattice}(b)] as follows.
In Fig.~\ref{fig_cv}(f), we show the temperature dependence of the flux density, which is computed for the site-diluted system by $\bar{W}=\frac{1}{L^2}\sum_p n_p  \means{W_p}$; 
$W_p$ is defined for all the plaquettes including larger ones than the six-sites hexagon and $n_p$ stands for the number of the original hexagons included in the plaquette $p$.
The result indicates that $\bar{W}$ is largely suppressed by the site dilution.
To reveal the origin of this behavior, we compute the average of $W_p$ only for the six-site plaquettes with $n_p=1$ remaining on the site-diluted lattice. 
As shown by the dashed lines in Fig.~\ref{fig_cv}(f), this quantity remains almost the same as in the pristine case, indicating that all $W_p=+1$ for the $n_p=1$ plaquettes in the low-temperature limit.  
Thus, the suppression of $\bar{W}$ is ascribed to fluctuations of $W_p$ for the plaquettes with larger $n_p$.
This is supported by considering the limit of $\rho \to 0$, where each vacancy yields a 12-site plaquette with $n_p=3$.
When we assume $\langle W_p \rangle=0$ for the $n_p=3$ plaquettes and $\langle W_p \rangle=1$ for the others with $n_p=1$, $\bar{W}$ should be $1-6\rho$, which well explains the low-temperature values of $\bar{W}$ in Fig.~\ref{fig_cv}(f) [see also Fig.~\ref{fig_disorder}(d)].
Thus, we conclude that the residual entropy in Fig.~\ref{fig_cv}(e) originates from the residual fluctuations of the fluxes in larger plaquettes yielded by vacancies. 
This appears to be consistent with the small flux-binding energy $\sim 0.003J$ for an isolated vacancy~\cite{Willans2010}.

\begin{figure}[t]
\begin{center}
  \includegraphics[width=\columnwidth,clip]{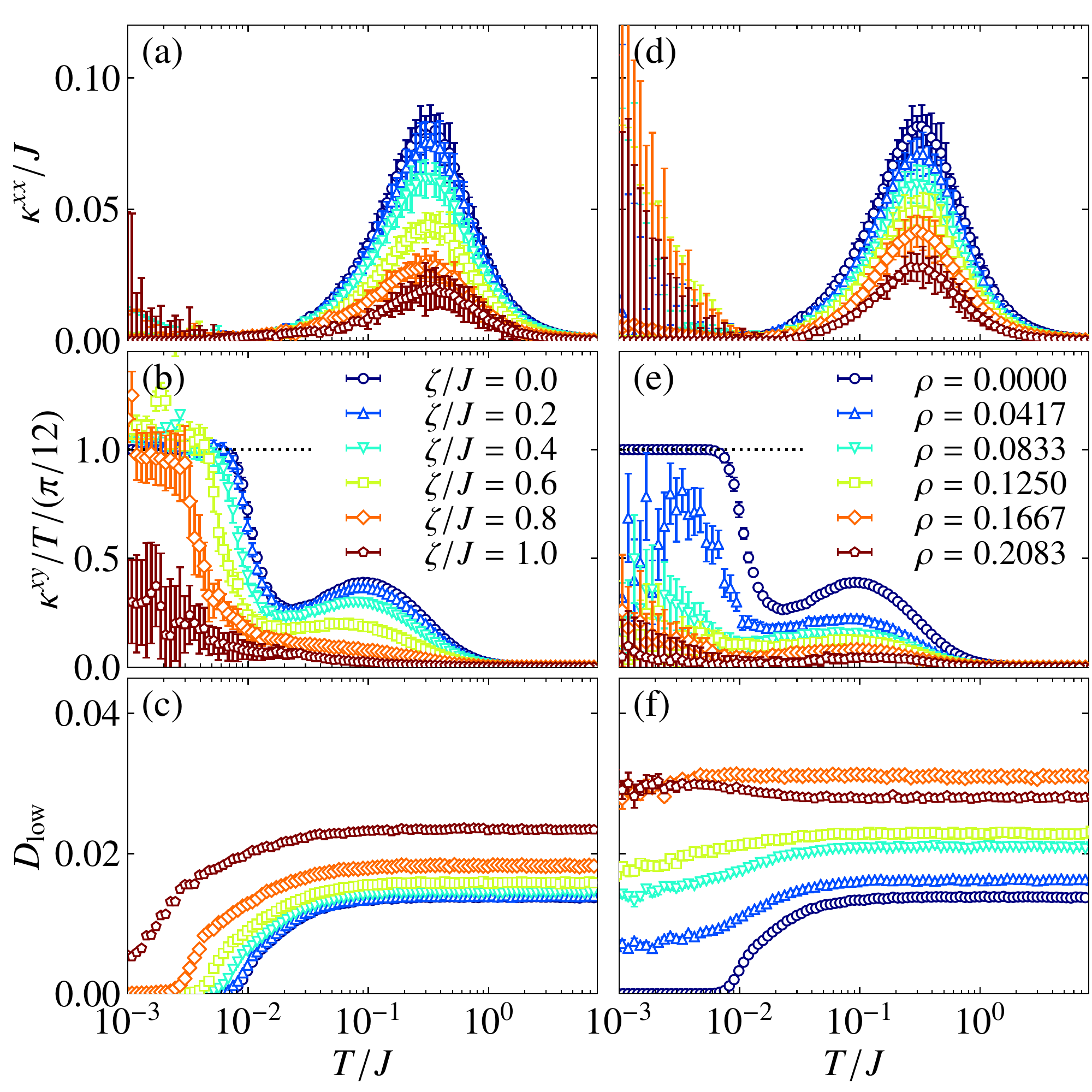}
  \caption{
    (a)--(c) Temperature dependences of (a) the longitudinal thermal conductivity, (b) the thermal Hall conductivity divided by temperature, and (c) the low-energy weight of the Majorana DOS in $0<\omega/J\leq 0.02$ for the systems with bond randomness.
    In (a), no magnetic field is applied but the data for (b) and (c) are the results under the effective magnetic field $\tilde{h}/J=0.06$.
    (d)--(f) Corresponding results for the systems with site dilution.
  }
  \label{fig_kappa}
\end{center}
\end{figure}

Now we turn to the thermal transport properties.
First, we discuss the longitudinal component of the thermal conductivity $\kappa^{xx}$, which is calculated in the same manner as Ref.~\cite{Nasu2017} (see also the footnote~\footnote{{The calculations both for the longitudinal and transverse thermal conductivities are carried out for the $10\times 10$ superlattice of the $N=288$ cluster.}}).
Figures~\ref{fig_kappa}(a) and \ref{fig_kappa}(d) show $\kappa^{xx}$ for the bond randomness and site dilution, respectively~\footnote{
  The errorbars include the errors from the extrapolation of $\kappa^{xx}(\omega)$ to $\omega\to 0$.
}.
In the pristine case, $\kappa^{xx}$ exhibits a broad peak around $T_H$, which indicates heat conduction by itinerant Majorana fermions~\cite{Nasu2017}. 
When introducing disorders, the peak is suppressed by both types of disorders.
In Figs.~\ref{fig_disorder}(a) and \ref{fig_disorder}(c), we plot the disorder dependence of $\kappa^{xx}$ around the peak temperature, in comparison with $C_v$ and the kinetic energy of the itinerant Majorana fermions per bond, $K$~\cite{PhysRevB.92.115122}. 
We find that while $C_v$ and $K$ do not change largely for both disorders, $\kappa^{xx}$ is strongly suppressed.
This suggests that the suppression of $\kappa^{xx}$ is caused by the reduction of the mean-free path $l\propto\kappa^{xx}/(v C_v)$ by the disorders, assuming that $K$ gives a measure of the velocity $v$ of the itinerant Majorana fermions.

Next, we examine the thermal Hall conductivity $\kappa^{xy}$ in a magnetic field.
Following Ref.~\cite{Nasu2017}, we compute $\kappa^{xy}$ in the presence of the effective magnetic field~\cite{Kitaev2006} by adding ${\cal H}_h= -\sum_{[jj''j']_{\alpha\beta\gamma}} \tilde{h}_{jj''j'} S_j^\alpha S_{j''}^\beta S_{j'}^\gamma$ to Eq.~(\ref{eq:Kitaev}), where $[jj''j']_{\alpha\beta\gamma}$ stands for neighboring three sites; 
the neighboring pair $jj''$ ($j''j'$) are located on $\alpha$ ($\gamma$) bond and $\beta$ is the component neither $\alpha$ nor $\gamma$.
For simplicity, the effective field is taken to be uniform as $\tilde{h}_{jj''j'}=\tilde{h}$, but in the case of the site dilution, it is set to zero if any of involved sites $j$, $j'$, and $j''$ is vacant.
Figures~\ref{fig_kappa}(b) and \ref{fig_kappa}(e) show $\kappa^{xy}/T$ at $\tilde{h}/J=0.06$ for the two types of disorders.
In the absence of disorder, $\kappa^{xy}/T$ approaches the quantized value $\pi/12$ below $T_L$~\footnote{We take the reduced Planck constant $\hbar=1$ and the Boltzmann constant $k_B=1$.}, reflecting the formation of the topological chiral QSL state under the magnetic field~\cite{Kitaev2006,Nasu2017}.
When introducing the disorder, the quantization plateau of $\kappa^{xy}/T$ shows contrasting responses to the two types of disorders.
In the bond-randomness case, although the onset temperature is reduced gradually while increasing $\zeta$, the quantization plateau remains for $\zeta/J \lesssim 0.8$ as shown in Fig.~\ref{fig_kappa}(c).
In contrast, it is fragile against the site dilution; 
it disappears even for $\rho \simeq 0.04$ in the calculated temperature range, as shown in Fig.~\ref{fig_kappa}(e) [see also Figs.~\ref{fig_disorder}(b) and \ref{fig_disorder}(d)].

Let us discuss the contrasting behavior of $\kappa^{xy}/T$ at low temperature% \nchange{in terms of the fractional excitations under the magnetic field}{}
. 
In the pristine case, the topological chiral QSL is realized by the flux condensation to the flux-free state with gap opening in the Majorana excitation~\cite{Kitaev2006}.
As shown in Figs.~\ref{fig_disorder}(b) and \ref{fig_disorder}(d), the disorder dependences of $\kappa^{xy}/T$ at low temperature are similar to those of $\bar{W}$ for both types of disorders.
This indicates the close relation between the quantization plateau and the flux condensation.
Note that, for the site dilution, the flux-free state is destroyed well below the percolation threshold $\rho_c\simeq 0.3$~\cite{Suding1999,Feng2008}.
In addition, we find that $\kappa^{xy}/T$ and $\bar{W}$ correlate with the Majorana excitation gap.
In the absence of disorder, the effective magnetic field opens a gap of $\frac{3\sqrt{3}}{4}\tilde{h}$ at $T=0$~\cite{Kitaev2006}, which is $\simeq 0.078J$ for $\tilde{h}/J=0.06$. 
The Majorana gap is perturbed in a different manner by the two types of disorders.
This is clearly demonstrated by calculating the low-energy weight of the Majorana density of states (DOS), $D_{\rm low}=\int_0^{\omega_c}D(\omega) d\omega$ by taking $\omega_c/J=0.02$ ($<0.078$).
As shown in Fig.~\ref{fig_kappa}(c), $D_{\rm low}$ remains almost zero below the onset temperature of the quantization plateau for the bond randomness.
In contrast, $D_{\rm low}$ becomes nonzero almost immediately by introducing the site dilution, as shown in Fig.~\ref{fig_kappa}(f). 
See also Figs.~\ref{fig_disorder}(b) and \ref{fig_disorder}(d). 
These contrasting effects on the fractional excitations underlie the contrasting behavior of the quantization plateau in Figs.~\ref{fig_kappa}(a) and \ref{fig_kappa}(d).

We discuss the relevance of our findings to experiments.
In (Na$_{1-x}$Li$_x$)$_2$IrO$_2$, which is regarded as a bond-disordered system, the peak of $C_v/T$ is suppressed and shifted to low temperature~\cite{Manni2014}. 
In H$_3$LiIr$_2$O$_3$ for which the effect of bond randomness was also discussed~\cite{Yadav2018,LiYing2018,Knolle2019disorder,Wang2018pre,Geirhos2020pre}, $C_v$ does not show any peak down to $0.05$~K~\cite{Kitagawa2018nature}. 
Our results in Fig.~\ref{fig_cv}(a) suggest a possible reinterpretation of these experiments from the fractional excitations in the presence of bond randomness, albeit subsidiary interactions beyond the Kitaev model are not taken into account. 
Meanwhile, in site-diluted systems $A_2$(Ir$_{1-x}$Ti$_x$)O$_3$ with $A$=Na and Li~\cite{Manni2014a} and (Ru$_{1-x}$Ir$_x$)Cl$_3$~\cite{Lampen-Kelley2017,Do2018,Do2020}, $C_v/T$ shows a hump shifting to low temperature by increasing $x$; 
interestingly, in the latter case, the hump remains at $\sim 3$~K where the magnetic order becomes vague~\cite{Do2018,Do2020}. 
This recalls the reduced hump in Fig.~\ref{fig_cv}(d), although the relation to the quantum spin glass was discussed~\cite{Georges2000,Georges2001,Camjayi2003,Takahashi2007,Watanabe2014,Kimchi2018,LiuLu2018,riedl2019critical,Andrade2014}.

Meanwhile, unusual contribution in $\kappa^{xx}$ was identified in $\alpha$-RuCl$_3$ and ascribed to itinerant Majorana fermions~\cite{hirobe2017}. 
Sample dependence was observed~\cite{Hentrich2018}, which might correspond to our results for $\kappa^{xx}$ in Figs.~\ref{fig_kappa}(a) and \ref{fig_kappa}(d).
Recent experiments on the thermal Hall conductivity also show sample dependence~\cite{Kasahara2018,kasahara2018majorana,Yokoi2020pre}.
A possible origin is the stacking fault, which may lead to bond randomness rather than site dilution. 
Our result in Fig.~\ref{fig_kappa}(b) suggests that the half quantization of $\kappa_{xy}/T$ is observed in high-quality samples with less stacking fault. 
The systematic decrease of the onset temperature accompanying the suppression of $\kappa^{xx}$ at high temperature, which are predicted in our results, would be worth testing in future experiments.

In summary, we have clarified that the two types of disorders, bond randomness and site dilution, have contrasting impacts on the thermodynamic and transport properties of the Kitaev model, through the fractional excitations of itinerant Majorana fermions and localized fluxes.
In particular, we found that the half-quantiztion of the thermal Hall conductivity is rather robust against the bond disorder but fragile for the site dilution.
Our results provide the systematic evolution of thermodynamics and thermal transport for disorders, which would be useful for identification of the Kitaev QSL in candidate materials and also the effects from other subsidiary interactions beyond the Kitaev model.

\begin{acknowledgments}
The authors thank M.~Shimozawa and Y.~Mizukami for fruitful discussions.
Parts of the numerical calculations were performed in the supercomputing
systems in ISSP, the University of Tokyo.
  This work was supported by Grant-in-Aid for Scientific Research from
  JSPS, KAKENHI Grant Nos. JP16H02206, JP18H04223, JP19K03742 and by JST PREST (JPMJPR19L5).
\end{acknowledgments}

\bibliography{refs}

\end{document}